%% file: KelvinGreenWriteup.tex
\documentclass[]{JFM-FLM_Au}

\usepackage{amsmath, amssymb, amsfonts}
\usepackage{hyperref}
\usepackage{graphicx}
\usepackage{svg}
\newcommand{\di}[1]{\mathop{\mathrm{d}#1}} 

\lefttitle{G. D. Weymouth}
\righttitle{Journal of Fluid Mechanics}
\title{Linear Kelvin Wave Predictions in the $z\to 0$ Limit}
\author{Gabriel D. Weymouth}
\corresau{\email{G.D.Weymouth@tudelft.nl}}
\affiliation{Ship Hydromechanics, Delft University of Technology, Delft, Netherlands}
\date{\today}

\begin{document}

\maketitle

\begin{abstract}
Linear wave theory captures the essential physics of free-surface flows at a fraction of the computational cost of nonlinear and viscous methods, making it attractive for design, real-time control, and surrogate modeling applications. However, the Kelvin Green's function for a translating point-source generates unbounded wave energy in the $z\to 0$ limit, causing both numerical difficulties and physical inconsistencies. This paper develops a modified kernel for the Kelvin potential incorporating an elliptic spanwise line integration that naturally resolves this ill-posedness, yielding finite wave energy over the entire free surface. We then present a fast evaluator for both point and line kernels using contour deformation adapted to the non-analytic Kelvin phase, achieving $10^4$-$10^5$ speedup over direct quadrature while preserving the wake asymptotics. Predictions on the most challenging $z=0$ limit demonstrate physically consistent wave patterns and wave resistance trends. An open-source Julia implementation is provided.
\end{abstract}


\section{Introduction}

Linear potential theory for steady forward-speed wave problems captures the essential physics of ship-wave interaction at a fraction of the computational cost of nonlinear and viscous flow models. This speed makes it attractive for applications such as design optimization, real-time control, and as the physics-informed backbone for machine learning surrogate models. However, the predictions also need to be free from empirical tuning parameters and robust for any possible input, which can be problematic for potential flows, as the absence of damping can lead to ill-posed or unbounded predictions.

The Kelvin Green's function for a steady translating disturbance exactly satisfies the linear free-surface boundary condition and has a well-established integral representation \citep{peters_new_1949}, but the potential is ill-posed as the source position approaches the free-surface. This singularity was first identified and analyzed by \cite{ursell_kelvins_1960}, and as shown in Section~\ref{sec:point_source_asymptotics}, the wave elevation spectrum $S_\zeta(k)$ peaks at $k^* \sim 1/|z|$ with amplitude $\sim 1/|z|^{3/2}$, diverging as $z\to 0$ and making the downstream wake unresolvable. A classic method to avoid the singularity is to introduce a damping term in the free-surface boundary condition \citep{havelock_theory_1932, furth_dissipative_2021}, but this complicates the Green's function evaluation and introduces a non-physical Rayleigh damping parameter.

Efficient and robust evaluation of the Green's function is challenging even for a well submerged source due to the oscillatory nature of the wavelike term's integrand, and this challenge intensifies as the exponential decay goes to zero with $z\to 0$, see Sec~\ref{sec:point_source_asymptotics}. This makes direct numerical quadrature feasible only at moderate submergence depths \citep{furth_dissipative_2021}.  Asymptotic series expansions have been developed, but require a patchwork of different series in different regions with no universally optimal series truncation criterion and unable to handle the $z\to 0$ limit for arbitrary distances from the source, \citep{baar_evaluation_1988, clarisse_evaluation_1994}. Numerical steepest-descent methods have been applied to the related problem of a pulsating translating Green's function \citep{iwashita_green_1992}, but the contour geometry used in that work degenerates in the steady limit. The steady problem has also been attempted with steepest-descent, but the exponential convergence mechanism used in steepest descent collapses as $z\to 0$ regardless \citep{motygin_numerical_2017}. Recent advances in contour deformation methods enable robust and efficient integration of analytic functions \citep{gibbs_numerical_2024}, but the Kelvin wavelike integral depends on a non-analytic phase function, preventing the direct application of these methods.

For surface-piercing bodies, the free-surface limit becomes unavoidable. The waterline contour contributes directly to the potential and must be evaluated at $z=0$ \citep{baar_developments_1988}. Classic flat ship theories were developed to predict the pressure on high-speed, shallow-draft vessels (with $z\approx 0$) but these are high Froude number approximations, not exact solutions, and do not automatically regularize the singular wave energy at the body waterline \citep{tuck_low-aspect-ratio_1975,cole_simple_1988}. The singular energy is integrable and labeled as ``spray drag'' by these methods, but the singularity remains unresolved. The persistent difficulty of evaluating the potential at $z=0$ motivated the development of the Neumann-Michell theory which uses an implicit iterative scheme for the potential to avoid the evaluation entirely \citep{noblesse_neumannmichell_2013}. However, this reformulation still uses an explicit truncation of the high wavenumber content to avoid the underlying singular behavior, which therefore remains unresolved.

This paper addresses this gap by establishing a well-posed, finite-energy formulation of the linear free-surface problem in the $z\to 0$ limit while retaining the explicit Green's function framework. The first contribution of this paper, after introducing the Kelvin Green's function and revisiting the divergence in the point-source spectrum in Section~\ref{sec:point_source_asymptotics}, is analytically developing the elliptic spanwise line kernel which naturally regularizes this divergence in Section~\ref{sec:line_integrated_kernels}. The second contribution is the development of a fast and robust numerical evaluator for both kernels in Section~\ref{sec:fast_evaluator} with an open-source Julia implementation \citep{weymouth_kelvinflatship_2026}. Finally, Section~\ref{sec:results} demonstrates the resulting wave fields and wave resistance of a rectangular flat-ship planform at $z=0$ across a range of Froude numbers.

\section{Kelvin Green's Function for Linear Potential Flow}

\begin{figure}
    \centering
    \def\svgwidth{0.75\textwidth}    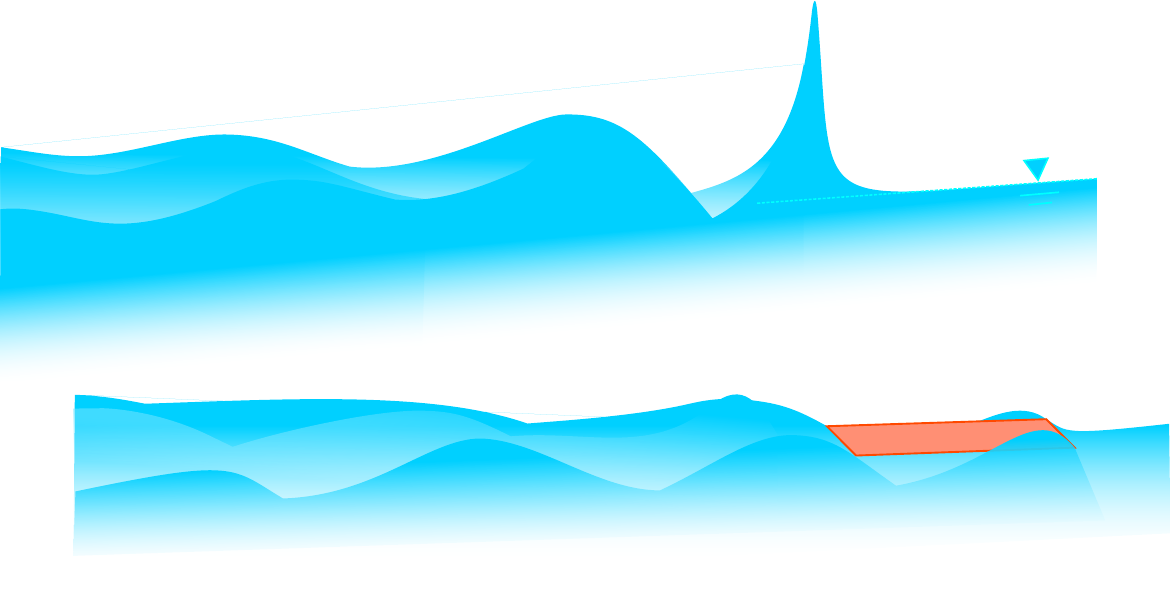
    \vspace{-3mm}
    \caption{Schematic of the free surface prediction problem with a submerged point source (top) and a flat-ship planform (bottom). The wave energy is unbounded in the wake of the point-source.}
    \label{fig:schematic}
\end{figure}

We consider steady, incompressible and irrotational flow in the lower half-domain $z\le0$ with a reference frame translating in direction $x$ at speed $\mathscr{U}$, Figure~\ref{fig:schematic}. After scaling all lengths by the Kelvin length $\ell=\mathscr{U}^2/\mathscr{g}$ (where $\mathscr{g}$ is the gravitational acceleration in direction $-z$) and the velocity potential $\phi$ by $\mathscr{U}\ell$, the potential must satisfy the linear free-surface boundary condition on the undisturbed free-surface plane
\[
\partial_{xx}\phi + \partial_z\phi = 0\quad\text{on}\quad z=0
\]
as well as $\nabla^2\phi = 0$ in the lower half-space.

The classic Kelvin Green's function of a translating point source at $\vec x_p$ satisfies the boundary condition above by construction and is made up of near-field terms $N$ which act close to the disturbance and a wavelike term $W$ whose influence extends into the wake
\[
G\bigl(\vec x_f,\vec x_p\bigr) = N(\vec x_f,\vec x_p) + W(\vec x_f - \vec x_p\!')
\]
where $\vec x_f$ is the field evaluation point and $\vec x_p\!'=(x_p,y_p,-z_p)$ is the image point reflected across $z=0$. As in \cite{noblesse_neumannmichell_2013}, we have combined the near-field terms together, and fast surrogates are known for their efficient evaluation, such as in \cite{newman_evaluation_1987}. In this work, we focus on the wavelike term.

Applying this Green's function, the induced potential of a source-patch $\mathcal{P}$ is
\begin{equation}\label{eq:patch}
\phi(\vec x_f) = \iint_{\mathcal{P}} q(\vec x_p)\,G(\vec x_f,\vec x_p)\,\di{a_p}
\end{equation}
where $q$ is the source strength scaled by $\mathscr{U}$. If $\mathcal{P}$ lies on or near $z=0$, then evaluating the linear free-surface elevation $\zeta = \partial_x\phi|_{z_f=0}$ and wave drag requires evaluating $G$, and in particular $W$, close to $z=0$. In the next section, we show that this limit is singular for a point source even far downstream. This divergence is well known since \cite{ursell_kelvins_1960} but we characterize it explicitly below as a basis for comparison.

\section{Point-Source Singularity as $z\to 0$}\label{sec:point_source_asymptotics}

We write the wavelike term in the general amplitude form
\begin{equation}\label{eq:wavelike_general}
W_A(x,y,z) = 4H(-x)\int_{-\infty}^{\infty} A(t)\,\exp\bigl(z(1+t^2)\bigr)\,\sin\bigl(g(x,y,t)\bigr)\,\di t
\end{equation}
where $\vec x=(x,y,z)$ is the relative vector from the image, $H$ is the Heaviside function, and the phase is $g = (x+yt)\sqrt{1+t^2}$. The variable of integration is $t=\tan\theta$ where $\theta$ is the wave propagation direction relative to $x$. The amplitude $A\equiv 1$ for a point-source, but this general form anticipates the analytic integration of Section~\ref{sec:line_integrated_kernels} and gives a unified form for numerical evaluation in Section~\ref{sec:fast_evaluator}.  

Stationary points of the wavelike phase satisfy $\partial_t g = 0$, and are given by
\begin{equation}\label{eq:sp}
t_\pm = -\frac{x \pm \sqrt{x^2 - 8y^2}}{4y}
\end{equation}
corresponding to the transverse and diverging wave systems inside the Kelvin wedge $|y| \le -x /\sqrt{8}$. Near the centerline we have $|y/x|\to 0$ and therefore the diverging ridge goes to $t_+ \approx -x/2y \to \infty$. The stationary-phase estimate for the diverging wave contribution to the potential on $z=0$ is
\[
W(x,y,0) \sim \frac{A_+}{\sqrt{\partial_{tt}g_+}}\sim \sqrt{\frac{t_+}{R}},\quad\text{where}\quad R = \sqrt{x^2+y^2}
\]
which grows without bound as $t_+\to\infty$ at finite $R$.

\begin{figure}
    \centering
    \includegraphics[width=0.55\linewidth]{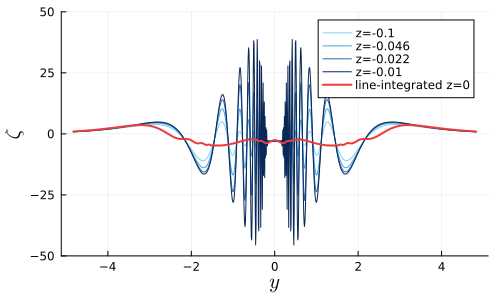}
    \includegraphics[width=0.44\linewidth]{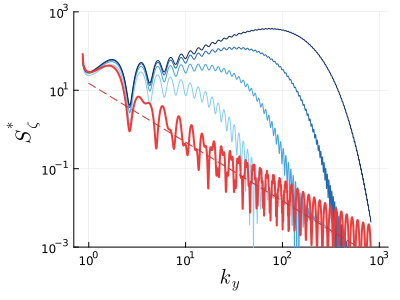}
    \caption{Free surface wave cut and spectra downstream of a unit point-source with $z\to 0$ and line-integrated source on $z=0$. The wave cuts on the left are taken at $x\mathbin{=}-8$. The spectra on the right are computed via Hilbert transformation for wave cuts at $x\mathbin{=}-40$. The modulations are due to transverse wave interference. The line-integrated results use half-beam $b\mathbin{=}1$ and $q_0\mathbin{=}\frac{2}{\pi}$ and the dashed line is the predicted $S_\zeta\sim k_y^{-3/2}$ decay.}
    \label{fig:wavecut}
\end{figure}

Exponential damping reduces the contribution of the diverging waves to the wavelike elevation spectrum $S_\zeta = |\partial_x W|^2$ when $z<0$, but the value is still unbounded in the limit. Defining 
\[k_x=\sqrt{1+t^2}\ ,\quad k_y=t k_x\ ,\quad k=|\vec k|=1+t^2\]
then for large $t_+$ we have $k_x\approx |t_+|$ and $|k_y|\approx k \approx t_+^2$, and stationary phase analysis on the wave elevation spectrum gives
\[
  S_\zeta(k) \sim \left(k_x\exp(-|z|k)\left(\frac{t_+}{R}\right)^{1/2}\right)^2 \sim \frac{k^{3/2}}{R}\,\exp\left(-2|z|k\right)
\]
where the $k_x$ factor comes from differentiation by $x$. This spectrum peaks at
\begin{equation}
  k^* = \frac{3}{4|z|}, \qquad S_\zeta(k^*) \sim \frac{1}{R\,|z|^{3/2}}.
  \label{eq:point_source_peak}
\end{equation}
As $z\to 0$, the peak migrates to an arbitrarily high wavenumber with unbounded amplitude. Critically, this is not a local artifact: the entire downstream wake is corrupted with non-physical wave-energy whose amplitude is only slowly modulated by $R^{-1}$. Figure~\ref{fig:wavecut} illustrates the unbounded growth as $z\to 0$ well downstream, in contrast with the regularized line-integrated kernel derived in the next section.

\section{Line-Integrated Potential for Flat-Ship Theory}\label{sec:line_integrated_kernels}

The unbounded wake energy identified in \eqref{eq:point_source_peak} is both nonphysical and numerically unresolvable, motivating the development of an integrated Kelvin Green's function for linear wave predictions of surface-piercing bodies. Approaches such as Neumann-Michell theory \citep{noblesse_neumannmichell_2013} avoid this difficulty by using integration by parts to reformulate the potential, but still require an explicit truncation of the wavelike spectrum, introducing a non-physical cutoff. The flat-ship case provides a uniquely challenging setting as the entire planform rests on $z=0$, Figure~\ref{fig:schematic}.

\subsection{The Flat-Ship Model}\label{ssec:flat_ship_model}

Consider an extremely shallow-draft ship such as a flat-bottom planing hull with a small pitch angle $\alpha$. In the limit $z_p\to 0^-$, the surface integral~\eqref{eq:patch} that defines the potential reduces via Green's theorem and the linear free-surface boundary condition to a waterline contour integral around the planform boundary $\partial\mathcal{P}$ as derived in \citep{noblesse_slender-ship_1983}
\[
\phi(\vec x_f) = \oint_{\partial\mathcal{P}} q(\vec x_p)\,G(\vec x_f-\vec x_p)\,n_x\,\di{y_p}.
\]
In Neumann-Kelvin methods \citep{baar_developments_1988}, this waterline contour is discretized with the goal of determining the source distribution $q$ by applying the body-boundary condition. However, until they are regularized, each of these integrals generates singular wave energy at unresolvable high wave number, making the inversion ill-posed. In contrast, the present work analytically evaluates the integral over a physically motivated trial strength distribution $q$, regularizing the integral and enabling universal finite energy predictions with an explicit Green's function.

For illustration, we focus on a rectangular planform of half-beam $b$ and length $L$.\footnote{As all lengths are scaled by $\ell=\mathscr{U}^2/\mathscr{g}$, the planform dimensions are inverse-square Froude-numbers; the length-based Froude number is $\mathrm{F_L}=L^{-1/2}$ and the half-beam-based Froude-number is $\mathrm{F_b}=b^{-1/2}$.} As $n_x = \pm 1$ on the leading and trailing edges and zero on the sides, the potential on the contour $\partial\mathcal{P}$ reduces to
\begin{equation}\label{eq:fs phi}
\phi(\vec x_f) = \Delta_L\int_{-b}^{b} q(x_p,y_p) G(x_f-x_p,y_f-y_p,z_f)\di{y_p}
\end{equation}
where $\Delta_L f(x_p) = f(0)-f(L)$ is the difference across the leading and trailing edges.

To determine the distribution of $q$, we impose a uniform downward velocity on the planform due to the small angle $\alpha$
\[
\partial_n\phi = \alpha, \qquad (x_p,y_p)\in\mathcal{P},\quad z_p\to 0^-.
\]
Substituting $G = N + W$ into \eqref{eq:fs phi}, and positing a solution where the high wavenumber components of $W$ are controlled, the dominant operator acting with $q$ to enforce the velocity condition is the derivative of the local logarithmic term:
\[
\int_{-b}^{b} q'(y_p)/(y-y_p)\,\di{y_p} = \mathrm{const},\qquad |y|<b.
\]
This is precisely the classical constant-downwash integral equation of finite-wing theory, whose solution on $[-b,b]$ vanishing at $y_p=\pm b$ is the elliptic distribution
\begin{equation}\label{eq:elliptic}
q(y_p) = q_0\sqrt{1-(y_p/b)^2}    
\end{equation}
with $q_0 \propto \alpha$. Using this elliptic distribution satisfies the body boundary condition to leading order and, as shown in the following, admits an exact Bessel function representation that regularizes the wavelike function for all $z$. As in classical thin-ship theory, this formulation neglects the wavelike influence on the hull. Corrections for the wave-interaction effects could in principle be incorporated with a Bessel function expansion or by building a Neumann-Kelvin approach from these regularized elements.

\subsection{The Line-Integrated Wavelike Kernel}

We define $W_b$ as the elliptically weighted spanwise integral of $W$
\begin{align*}
  W_b(x,y,z) &= \int_{-b}^{b} \sqrt{1-(y_p/b)^2}\quad\! W(x,y-y_p,z)\, \di{y_p} \notag\\
  &=  4H(-x)\int_{-\infty}^{\infty} \exp\left(z(1+t^2)\right) \int_{-b}^{b} \sqrt{1-\left(\frac{y_p}{b}\right)^2} \,\sin\bigl(g(x,y-y_p,t)\bigr)\,\di{y_p}\,\di t
\end{align*}
where we have swapped the order of the two integrals. The inner integral is the Fourier transform of the elliptic distribution, which has the exact Bessel function representation
\[
\int_{-1}^1 \sqrt{1-\eta^2} e^{i\omega\eta} \di\eta= \pi \frac{J_1(\omega)}\omega
\]
where $\mathrm{J}_1$ is the Bessel function of the first kind. Substituting $\eta=y_p/b$ and $\omega=bk_y$, the wavelike kernel $W_b$ takes the general form~\eqref{eq:wavelike_general} with amplitude
\begin{equation}
  A_b(t) = \pi\,\frac{\mathrm{J}_1\bigl(b\,k_y(t)\bigr)}{k_y(t)}.
  \label{eq:Ab}
\end{equation}
The amplitude function $A_b$ is finite at $k_y=0$ (with limit $\frac 12 \pi b$) and decays as $k_y^{-3/2}$ for $k_y \gg 1/b$, filtering high wavenumber content even on $z=0$. Applying the stationary-phase analysis of Section~\ref{sec:point_source_asymptotics} to $W_b|_{z=0}$ gives the wave elevation spectrum
\[
  S_{b,\zeta}(k) \sim \frac{\mathrm{J}_1^2\left(b\,k_y\right)}{k_y^2}\frac{k_y^{3/2}}{R} \sim \frac{1}{k_y^{3/2}bR} 
\]
which is uniformly decaying and integrable, completely eliminating the unbounded high-wavenumber energy of the point-source kernel. This is demonstrated in Figure~\ref{fig:wavecut} for $b=1$. Note that the wave energy remains finite for any $b>0$ since the point-source singularity is recovered only if the transition wavenumber $k_y\sim 1/b$ diverges.

\subsection{Wave Resistance}\label{ssec:wave_resistance}

The wave spectrum leads directly to the flat-ship wave resistance $\mathscr{D}_W$ from the classic \cite{havelock_theory_1932} formula. Using $k_x= \sqrt{1+t^2}=\sec\theta$ and wake symmetry we have
\begin{equation}
  \mathscr{D}_W = \frac{8\pi\rho}{\ell^2} \int_{-\pi/2}^{\pi/2} |\mathscr{H}(\theta)|^2\,\sec^3\theta\,\di\theta = \frac{16\pi\rho}{\ell^2} \int_0^{\infty} |\mathscr{H}(t)|^2\, k_x(t)\,\di t
  \label{eq:havelock}
\end{equation}
where $\rho$ is the fluid density and $|\mathscr{H}(t)|$ is the amplitude of the wavelike integral over the planform. For the elliptic spanwise distribution and rectangular planform we have
\[
\mathscr{H}(t) = q_0\mathscr{U}\ell^2\,A_b(t)\Delta_Le^{ix_pk_x}.
\]
Substituting into~\eqref{eq:havelock} and defining a resistance coefficient $C_W = \mathscr{D}_W/(\rho \mathscr{U}^2 (2b\ell)^2)$ gives
\begin{equation}
  C_W = 8\pi q_0^2 \int_0^{\infty}
  A_w\bigl(1 - \cos\bigl(Lk_x\bigr)\bigr)\,\di t,\qquad A_w=\left(\pi\frac{\mathrm{J}_1(b\,k_y)}{bk_y}\right)^2 k_x.
  \label{eq:Cw}
\end{equation}
The amplitude $A_w$ is $O(1)$ near $t\approx0$ and decays as $t^{-5}$ for large $t$, so~\eqref{eq:Cw} is well-posed and rapidly convergent.

\section{Partitioned evaluation method for wavelike kernels}\label{sec:fast_evaluator}

Direct numerical quadrature and series expansions for the wavelike integral \eqref{eq:wavelike_general} are inefficient or even non-convergent as $z\to0$. In this section, we detail an efficient and universally robust partitioned contour-deformation approach for the wavelike kernels, leveraging the classical stationary phase results above. An open-source Julia implementation is provided at \cite{weymouth_kelvinflatship_2026}.

\subsection{Partition boundaries and quadrature methods}

\begin{figure}
    \centering
    \includegraphics[width=0.495\linewidth]{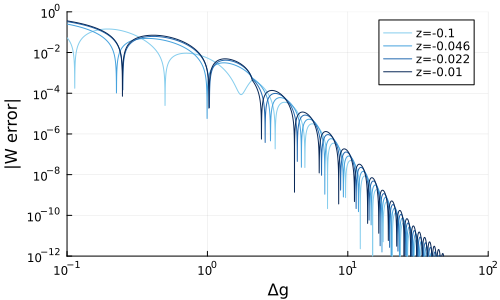}
    \includegraphics[width=0.495\linewidth]{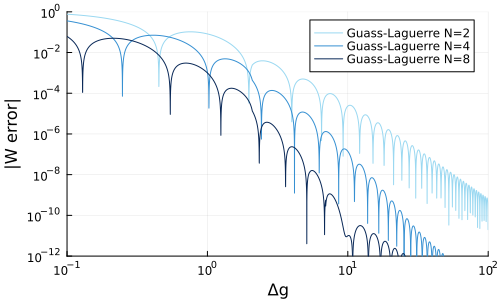}
    \caption{Error convergence of the partitioned quadrature method applied to $W(-8,y,z)$ on the same wavecuts in figure \ref{fig:wavecut}. Left shows the independence of the error as $z\to 0$ for $N=4$ Gauss-Laguerre points, and right shows the trend with $N$ for $z=-0.01$.}
    \label{fig:delta_phase}
\end{figure}

As in \cite{gibbs_numerical_2024}, the wavelike integral is partitioned into two types of regions; (i) non-oscillatory intervals around stationary points, where standard quadrature is efficient, and (ii) the remaining semi-infinite tails, where complex contour quadratures converge exponentially. 

Unlike the analytic phase functions treated in Gibbs et al., the wavelike phase $g(x,y,t) = (x+yt)\sqrt{1+t^2}$ is non-analytic and introduces branch points at $t = \pm i$. Therefore, we limit the non-oscillatory intervals to the real axis, defined such that each interval endpoint $h$ satisfies
\[
|g(x,y,h) - g(x,y,t_*)| = \Delta g
\]
where $t_*\in t_\pm$ are the stationary points~\eqref{eq:sp} and $\Delta g$ is a prescribed phase increment. Limiting the 
intervals to the real axis also simplifies locating the endpoints $h$ with numerical 1D root-finding. Overlapping non-oscillatory intervals are merged, naturally handling the coalescence of the transverse and diverging wave ridges as $y\to \pm x/\sqrt{8}$. Outside this wedge, a single pseudo-stationary point $t_*=-x/4y=\Re(t_\pm)$ is used, ensuring predictive continuity across the Kelvin wedge boundary. 

The kernel is smooth over the non-oscillatory interval by construction, meaning direct quadrature can be applied efficiently on the real line. For this manuscript, we use adaptive Gauss-Kronrod quadrature to ensure a prescribed accuracy. Each semi-infinite tail contribution to integral~\eqref{eq:wavelike_general} is written as
\[
 \pm\,\Im\left(\int_h^{\pm\infty} A(t) e^{z(1+t^2)+ig(x,y,t)}\di t \right) 
\]
where the complex phase combines the oscillatory and exponential terms. Each tail is evaluated using numerical steepest descent, with contour points in the complex plane located by Newton's method and Gauss-Laguerre quadrature. The $\sqrt{1+t^2}$ branch is selected such that the phase remains continuous along each contour.

Figure~\ref{fig:delta_phase} shows the error of the point-source wavelike integral using this partitioned quadrature method. The error envelope is insensitive to $\vec x$, handling the wave caustic and the energy growth as $z\to 0$ gracefully. The partition half-width $\Delta g$ is the primary numerical parameter of the approach, encoding a tradeoff between the number of oscillations along the real line and the exponential decay along the contours. If the contour starts too close to a stationary point (small $\Delta g$), the phase is slowly varying and the integrand decays slowly in the complex plane. Conversely, large $\Delta g$ ensures rapid decay and efficient contour quadrature, but increases the oscillations in the real-axis intervals, making them more expensive to integrate. The number of Gauss-Laguerre points $N$, is a secondary parameter: increasing $N$ increases the error decay rate, but this is only significant when $\Delta g$ is large, and it increases the length of the contours and computational expense. Herein, we set $\Delta g \approx 2\pi$ and $N=4$, producing maximum errors around $10^{-6}$ with minimal computational cost. In comparison, methods based purely on steepest-descent such as \cite{motygin_numerical_2017} lose their exponential convergence as $z\to 0$, requiring a rapid increase in the number of function evaluations in this limit. To reduce the error to $10^{-6}$, for example, that method requires between $10^5-10^6$ function evaluations for points near $z=0$. The current partitioned approach uses only $O(10)$ evaluations for the complex tails and another $O(10)$ real-line evaluations, leading to evaluation times on the order of $1\mu s$, Table~\ref{tab:wavelike}.

\subsection{Bessel function decomposition}

\begin{table}
    \centering
    \begin{tabular}{r|ccccc}
    $y$    & absolute error   & relative error  & time ($\mu s)$ & speedup   & slowdown \\[1pt]
    \midrule[0.5pt]\\[-8pt]
    0    & 3.78e-6 & 2.64e-7 &  30.4 & 6.59e4 & 7.02   \\
    0.5  & 1.16e-6 & 9.45e-8 &  34.3 &  1.77e5 & 6.45  \\
    0.9  & 1.14e-6 & 1.56e-7 & 131.7 &  7.86e4 & 26.6  \\
    1.1  & 2.09e-6 & 4.14e-7 & 149.5 &  1.20e5 & 29.4  \\
    1.35 & 2.75e-6 & 9.14e-7 &  57.4 &  6.22e5 & 11.6
    \end{tabular}
    \caption{Error and timing of the partitioned quadrature approach applied to $W_b(-1,y,0)$ for $b=1$. Times measured on an Intel i9 laptop. Error and speed up are relative to an optimized adaptive Gauss-Kronrod quadrature method applied to $W_b$. Slowdown is relative to the partitioned quadrature applied to the point-source integral $W(-1,y,0)$.}
    \label{tab:wavelike}
\end{table}

The same contour-deformation strategy applies to the line-integrated source, extended to handle the additional oscillatory structure introduced by the Bessel function in $A_b$ \eqref{eq:Ab}. To isolate these oscillations, the Bessel function is decomposed using Hankel functions
\[
\mathrm{J}_1(\omega) = \frac{1}{2}\left(\mathrm{Hx}_1^+(\omega)\,e^{i\omega} + \mathrm{Hx}_1^-(\omega)\,e^{-i\omega}\right)
\]
where the exponentially scaled Hankel functions $\mathrm{Hx}_1^\pm$ are slowly varying. The exponential factors oscillate at a frequency $\omega = bk_y$, but these can be absorbed into the complex phase, yielding two integrals with shifted phases $g(x,y \pm b,t)$. Each shifted phase has its own stationary points and associated non-oscillatory interval. As before, any overlapping intervals are merged and integrated along the real line, now using the $A_b$ amplitude. The semi-infinite ranges are computed once with each phase and corresponding amplitude
\[
A_b^\pm = \frac{\mathrm{Hx}^\pm_1(b k_y)}{2k_y}
\]
using the identity $\mathrm{Hx}^\pm_1(\omega) = -\mathrm{Hx}^\mp_1(-\omega)$ as needed to avoid the $\Re(\omega)<0$ branch cut.\footnote{The same approach can also be applied to the wave resistance integral~\eqref{eq:Cw}. However, the faster $A_w\sim t^{-5}$ decay and slower $k_x$ phase function mean direct quadrature is practical for that integral.}

In general, the $W_b$ quadrature requires approximately twice as many phase and kernel evaluations as $W$, and Bessel function evaluations are $O(10)$ times as expensive as simple trigonometric functions. However, the computational cost is still more than $10^4$--$10^5$ times faster than direct quadrature on $z=0$ with relative errors less than $10^{-6}$, Table~\ref{tab:wavelike}.

\section{Flat-ship wave predictions}\label{sec:results}

\begin{figure}
    \centering
    \includegraphics[width=0.485\linewidth,trim=0 0 60 0,clip]{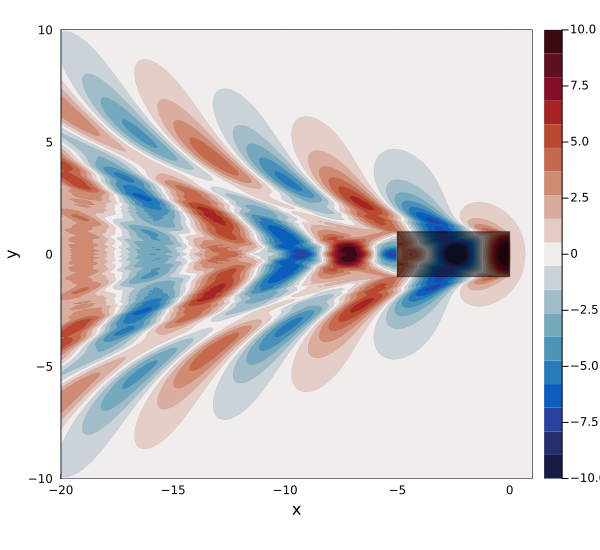}
    \hfill
    \begin{minipage}[]{0.445\linewidth}
        \vspace{-65mm}
        \includegraphics[width=\linewidth,trim=30 291 60 0,clip]{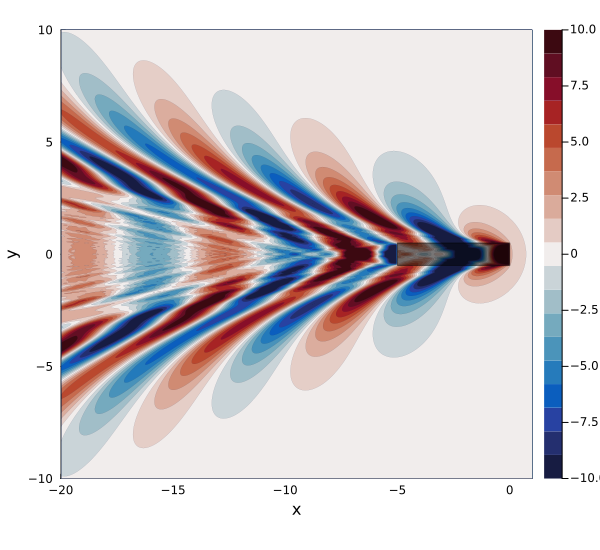}\\[1pt]
        \includegraphics[width=\linewidth,trim=30 0 60 249,clip]{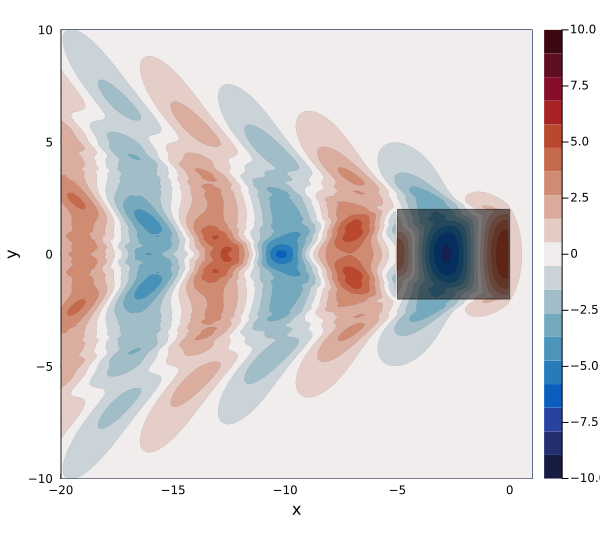}
    \end{minipage}
    \hfill
    \includegraphics[width=0.05\linewidth,trim=545 0 0 0,clip]{figures/wave_field_b1.png}
    \caption{Flat-ship wave elevation $\zeta$ scaled by the spanwise integrated strength $\tfrac \pi 2 q_0b$ for $L=5$. Left: full symmetric field for $b=1$. Right: half-field for $b=1/2$ (top) and $b=2$ (bottom) using the same contour levels.}
    \label{fig:wave_elevation}
\end{figure}

Using the elliptic distribution, the flat-ship potential \eqref{eq:fs phi} becomes
\[
\phi(\vec x_f) = q_0\Delta_L\left[ W_b(x_f-x_p,y_f,z_f)+\int_{-b}^b \sqrt{1-\left(\frac{y_p}b\right)^2}\, N(x_f-x_p,y_f-y_p,z_f)\,\di{y_p} \right]
\]
where the near-field term $N$ can be numerically integrated without issue and the wavelike term is calculated using the partitioned contour deformation method described in Section~\ref{sec:fast_evaluator}. Automatic Differentiation is used to evaluate the $\zeta=\partial_x\phi$ derivative.

Figure~\ref{fig:wave_elevation} shows a set of resulting flat-ship wave-fields for various $b$. First, we note that the waves predicted using the elliptic distribution on $z=0$ have finite amplitude everywhere. Since submergence adds exponential wave damping, achieving finite energy on $z=0$ guarantees well-behaved predictions for any submergence. In addition, the potential retains the expected logarithmic singularity across the line-source, resulting in a finite jump in $\zeta$ across the planform bow and stern edges. This discontinuity represents the pressure jump across the planform boundary and does not generate singularities downstream.

Second, we note that each corner of the planform generates a distinct wave train, with the theory analytically capturing the strong dependence of these waves on the finite planform width $b$. Directly behind the planform the interaction of the stern corner waves generates a large rooster-tail in the near wake, leaving a rough but low amplitude wave-field in the far wake after the wave trains diverge. The theory predicts a strong decrease in the diverging wave elevation amplitude with increasing $b$ for a given integrated line-source strength $\frac \pi 2 q_0 b$. Intuitively, increasing the Kelvin-scaled beam $b$ takes the planform out of strong resonance with the Kelvin wavelengths $\lambda=2\pi\ell\cos^2\theta$. More specifically, as $b$ increases, the spanwise edges generate diverging waves with increasingly diverse spanwise phases, increasing the destructive interference. This mechanism is encoded in the Bessel function amplitude $A_b$~\eqref{eq:Ab} which filters high-wavenumber contributions more aggressively as $b$ increases. In contrast, the transverse wave has $k_y\approx0$ where $A_b \approx \frac \pi 2 b$, resulting in a roughly constant transverse amplitude after scaling by $\frac \pi 2 q_0b$.

\begin{figure}
    \centering
    \includegraphics[width=0.9\linewidth]{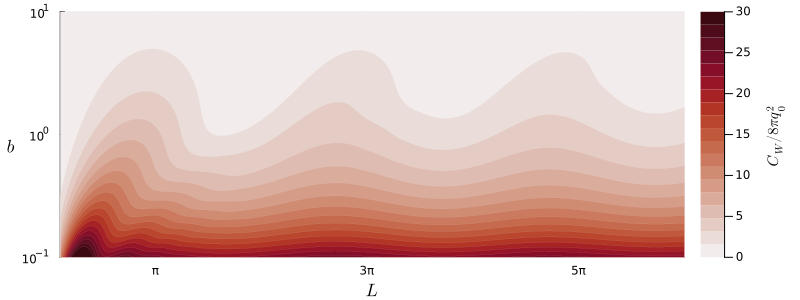}
    \caption{Flat ship wave resistance coefficient $C_W$ from \eqref{eq:Cw} scaled by $8\pi q_0^2$ as a function of the Kelvin-scaled planform length $L=\mathrm{F}_L^{-2}$ and width $b=\mathrm{F}_b^{-2}$.}
    \label{fig:wave_drag}
\end{figure}

Figure~\ref{fig:wave_drag} shows scaled resistance~\eqref{eq:Cw} across a wide range of $b,L$. The interference factor $1-\cos(Lk_x)$ produces classical constructive and destructive interference between the bow and stern wave systems, resulting in oscillating resistance values with $L$. More uniquely, the destructive spanwise interference discussed above results in a scaling $A_w\sim b^{-3}$ for large $b$, causing a gradual but monotonic decrease in the resistance coefficient $C_W$ with increasing $b$. This contrasts with classical thin-ship theory, which uses the beam-to-length ratio as a small parameter and predicts that wave resistance increases quadratically with beam, resulting in a constant $C_W$ regardless of $b$. The new flat-ship theory captures the nonlinear variation in the potential with $b$, encoded in the Bessel function of the wavelike kernel, resulting in the drop in $C_W$ due to destructive diverging wave interference missing in thin-ship theory.

\section{Conclusions}

The singular behavior of the point-source Kelvin Green's function as $z\to 0$ is a fundamental inconsistency which produces unbounded wave energy even far downstream, but the present work shows that this pathology is not intrinsic to linear wave theory. An elliptic source distribution, motivated by the finite disturbance width, yields a regularized kernel with finite energy and consistent spectral decay, without the introduction of empirical damping, arbitrary wavenumber filtering, or implicit formulations. 

The resulting formulation produces well-behaved free-surface predictions directly on $z=0$, capturing Froude number trends and wave interference patterns consistent with finite-width ships. The $k_y^{-3/2}$ spectral decay of the regularized kernel ensures finite wave energy and elevation everywhere, but higher-order derivatives of the wave field will eventually diverge at small scales, due to the absence of nonlinear and viscous dissipative mechanisms. However, the present formulation resolves the ill-posedness in the physically relevant quantities of wave elevation, energy, and drag within the inviscid linear framework. 

The present work focuses on the leading-order flat-ship approximation, neglecting nonlocal wave interactions on a finite depth hull. Accounting for local hull geometry while retaining the regularized kernel framework is a natural extension, at which point the predictions could be compared to experiments or high-fidelity nonlinear simulations. However, the contour-deformation approach developed here is completely general, evaluating both point and line kernels efficiently (achieving $10^4$--$10^5$ speedup over direct quadrature) while maintaining accuracy across the entire lower-half domain. Therefore the regularized kernel and efficient evaluator form a foundation for a well-posed Kelvin solver fast enough for use in applications requiring repeated evaluations, including design optimization and physics-informed surrogate modeling.

\bibliographystyle{jfm}
\bibliography{references}
\end{document}

%% file: figures/flat_ship_schematic.pdf_tex
\begingroup%
  \makeatletter%
  \providecommand\color[2][]{%
    \errmessage{(Inkscape) Color is used for the text in Inkscape, but the package 'color.sty' is not loaded}%
    \renewcommand\color[2][]{}%
  }%
  \providecommand\transparent[1]{%
    \errmessage{(Inkscape) Transparency is used (non-zero) for the text in Inkscape, but the package 'transparent.sty' is not loaded}%
    \renewcommand\transparent[1]{}%
  }%
  \providecommand\rotatebox[2]{#2}%
  \newcommand*\fsize{\dimexpr\f@size pt\relax}%
  \newcommand*\lineheight[1]{\fontsize{\fsize}{#1\fsize}\selectfont}%
  \ifx\svgwidth\undefined%
    \setlength{\unitlength}{561.79768125bp}%
    \ifx\svgscale\undefined%
      \relax%
    \else%
      \setlength{\unitlength}{\unitlength * \real{\svgscale}}%
    \fi%
  \else%
    \setlength{\unitlength}{\svgwidth}%
  \fi%
  \global\let\svgwidth\undefined%
  \global\let\svgscale\undefined%
  \makeatother%
  \begin{picture}(1,0.50322465)%
    \lineheight{1}%
    \setlength\tabcolsep{0pt}%
    \put(0,0){\includegraphics[width=\unitlength,page=1]{flat_ship_schematic.pdf}}%
    \put(0.80679146,0.14664695){\rotatebox{2.5880748}{\makebox(0,0)[t]{\lineheight{1.25}\smash{\begin{tabular}[t]{c}$L$\end{tabular}}}}}%
    \put(0.76949929,0.08718595){\color[rgb]{1,0.24313725,0}\rotatebox{0.10652211}{\makebox(0,0)[t]{\lineheight{1.25}\smash{\begin{tabular}[t]{c}$\cal P$\end{tabular}}}}}%
    \put(0.7263117,0.12458986){\rotatebox{-50.49668833}{\makebox(0,0)[t]{\lineheight{1.25}\smash{\begin{tabular}[t]{c}$2b$\end{tabular}}}}}%
    \put(0,0){\includegraphics[width=\unitlength,page=2]{flat_ship_schematic.pdf}}%
    \put(0.75449121,0.23328259){\color[rgb]{0,0,0}\makebox(0,0)[lt]{\lineheight{1.25}\smash{\begin{tabular}[t]{l}$\mathscr{U}$\end{tabular}}}}%
    \put(0,0){\includegraphics[width=\unitlength,page=3]{flat_ship_schematic.pdf}}%
    \put(0.96029817,0.28366269){\color[rgb]{0,0,0}\transparent{0.91372502}\makebox(0,0)[lt]{\lineheight{1.25}\smash{\begin{tabular}[t]{l}$\mathscr{g}$\end{tabular}}}}%
    \put(0.70616192,0.2755686){\color[rgb]{0,0,1}\transparent{0.91372502}\makebox(0,0)[lt]{\lineheight{1.25}\smash{\begin{tabular}[t]{l}$\vec x_p$\end{tabular}}}}%
    \put(0.70616192,0.39015676){\color[rgb]{0,0,1}\transparent{0.63960749}\makebox(0,0)[lt]{\lineheight{1.25}\smash{\begin{tabular}[t]{l}$\vec x_p\!'$\end{tabular}}}}%
    \put(0,0){\includegraphics[width=\unitlength,page=4]{flat_ship_schematic.pdf}}%
    \put(0.39762759,0.23122852){\color[rgb]{0,0,0}\transparent{0.91372502}\makebox(0,0)[lt]{\lineheight{1.25}\smash{\begin{tabular}[t]{l}$\vec x_f$\end{tabular}}}}%
    \put(0.55592997,0.32548192){\color[rgb]{0,0,0}\transparent{0.91372502}\rotatebox{31.17183135}{\makebox(0,0)[t]{\lineheight{1.25}\smash{\begin{tabular}[t]{c}$\vec x$\end{tabular}}}}}%
    \put(0,0){\includegraphics[width=\unitlength,page=5]{flat_ship_schematic.pdf}}%
  \end{picture}%
\endgroup%